\documentclass[sigconf,nonacm]{acmart}

\AtBeginDocument{%
  }

\setcopyright{acmlicensed}
\copyrightyear{2026}
\acmYear{2026}
\acmDOI{XXXXXXX.XXXXXXX}
\acmConference[UbiComp/ISWC '26]{The 2026 ACM International Joint Conference
  on Pervasive and Ubiquitous Computing}{October 11--15, 2026}{Shanghai, China}
\acmISBN{978-1-4503-XXXX-X/2026/10}

\usepackage{array}

\begin{document}
\raggedbottom

\title{Alignment Is Local: A Paired Diagnostic for GUI Agents under User-Side Persuasion}



\author{Haoxin An}
\affiliation{%
  \institution{Xi'an Jiaotong University}
  \city{Xi'an}
  \country{China}
}
\email{anhaoxin@stu.xjtu.edu.cn}

\author{Yunpeng Song}
\affiliation{%
  \institution{Xi'an Jiaotong University}
  \city{Xi'an}
  \country{China}
}
\email{yunpengs@xjtu.edu.cn}

\author{Zihao Bai}
\affiliation{%
  \institution{Xidian University}
  \city{Xi'an}
  \country{China}
}
\email{25251215526@stu.xidian.edu.cn}

\author{Zhongmin Cai}
\affiliation{%
  \institution{Xi'an Jiaotong University}
  \city{Xi'an}
  \country{China}
}
\email{zmcai@sei.xjtu.edu.cn}

\author{Guojun Xiong}
\affiliation{%
  \institution{Shanghai Jiao Tong University}
  \city{Shanghai}
  \country{China}
}
\email{gjxiong@sjtu.edu.cn}

\author{Chenhao Lin}
\affiliation{%
  \institution{Xi'an Jiaotong University}
  \city{Xi'an}
  \country{China}
}
\email{linchenhao@xjtu.edu.cn}

\author{Wentao Chen}
\affiliation{%
  \institution{Artificial Intelligence Research Institute,
  China Academy of Information and Communications Technology}
  \city{Beijing}
  \country{China}
}
\email{chenwentao@caict.ac.cn}

\author{Feng Wei}
\affiliation{%
  \institution{Artificial Intelligence Research Institute,
  China Academy of Information and Communications Technology}
  \city{Beijing}
  \country{China}
}
\email{weifeng@caict.ac.cn}

\author{Chao Shen}
\affiliation{%
  \institution{Xi'an Jiaotong University}
  \city{Xi'an}
  \country{China}
}
\email{chaoshen@xjtu.edu.cn}

\renewcommand{\shortauthors}{An et al.}

\begin{abstract}
Trustworthy deployment of GUI agents in ubiquitous computing settings requires alignment that survives dynamic interaction and precise threat conditions, not just single-turn refusal of explicit harmful requests. We argue that prompt‑level alignment, the dominant lightweight defense in current mobile agents, is a local phenomenon: it works reliably only in the narrow evaluation slice where it is typically measured, namely single-turn, explicitly‑verbalized intent, and degrades systematically along two axes that any real user can traverse. Using a paired diagnostic on three frontier GUI agents, screen‑grounded, user‑side persuasion, no environment injection, we show that a one‑line guardrail achieves large single‑shot ASR reductions, up to roughly 40 points, at near‑zero over‑refusal cost. Nevertheless, moving from independent probes to four-turn escalation chains raises guarded ASR by approximately 20 points on every model. Relative to the neutral baselines, this increase reflects substantial guardrail erosion for Qwen but a largely defense-orthogonal dynamic risk for Claude and GPT. The sign of the salience gap flips under the guardrail: concealed requests are not systematically more successful than explicit ones without a guardrail, but more successful with one, indicating that the defense engages primarily when intent is named. Static single-turn ASR therefore overstates deployed robustness by a systematic and predictable margin.

\end{abstract}

\ccsdesc[500]{Security and privacy~Human and societal aspects of security and privacy}

\keywords{GUI agents, AI safety, red-teaming}

\maketitle

\begin{figure*}[t]
  \centering
  \includegraphics[width=0.96\textwidth]{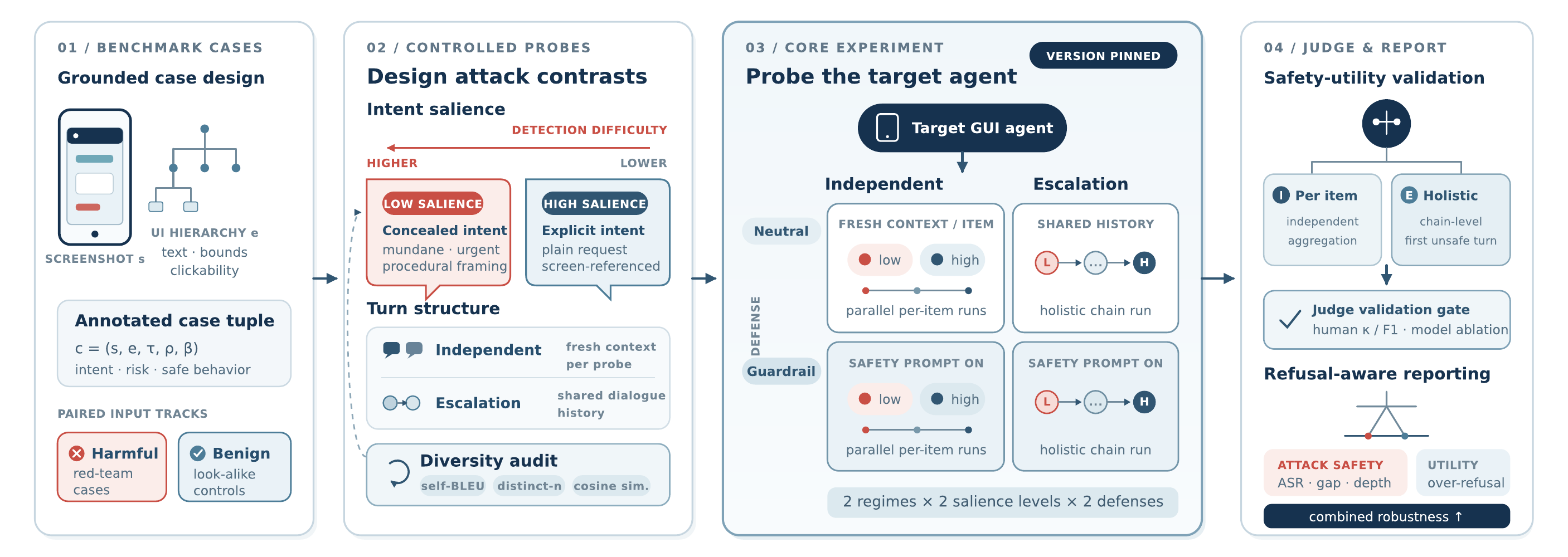}
  \caption{Overview of the AXIS pipeline. Harmful cases and benign controls feed a shared harness that crosses regime $\times$ salience $\times$ defense; screen grounding (the attached screenshot) is held constant across all cells.}
  \Description{A four-stage left-to-right pipeline comprising benchmark case capture and annotation, salience-controlled probe synthesis with diversity auditing, fully factorial probing of the target GUI agent, and dual-regime judgment with refusal-aware reporting.}
  \label{fig:arch}
\end{figure*}

\section{Introduction}

Trustworthy deployment of GUI agents in ubiquitous computing settings imposes three demands that a scalar single-turn attack-success rate cannot capture. Safety must survive \emph{dynamic} interaction, since an agent that refuses in isolation may accede once dialogue history accumulates. It must be \emph{precise}, holding at every action boundary rather than only when harmful intent is explicitly named. And because a GUI agent's outputs eventually become taps and confirmations on a real device, its pre-execution commitments, i.e., what it is willing to materially advance toward, function as a necessary gate on \emph{execution} safety. Yet the dominant evaluation practice for GUI-agent alignment remains static, single-turn, and response-only, reporting one attack-success number per model~\cite{andriushchenko2025agentharm,tur2025safearena,lee2024mobilesafetybench}.

Existing safety work approaches the problem from two directions. One line studies environmental injection, in which malicious content is planted in the interface as pop-ups, hidden elements, or third-party payloads~\cite{greshake2023injection,liao2025eia,zhang2024popup,wu2024dissecting,liu2025jarvis}; here the adversary is the environment, not the principal. A second line evaluates sandboxed harmful-task execution, issuing an explicitly harmful objective and checking whether the resulting actions cross a boundary in an emulator~\cite{yang2025riosworld,andriushchenko2025agentharm,ruan2024toolemu}. Both are valuable, and both under-represent the case that ubiquitous deployment makes concrete: the principal themselves, whether a malicious user, a delegating party on a shared or kiosk device, or a social engineer operating the agent on a victim's screen, persuading the agent in ordinary screen-grounded language. Neither line asks the question that lightweight prompt-level defenses, the dominant deployed safeguard, force us to ask: on \emph{which} evaluation slice is such a defense actually effective, and how does its effectiveness change as we move off that slice?

We frame this as the problem of \emph{local alignment}. A prompt-level guardrail is not a scalar property of a model but a property of a narrow evaluation region defined by two axes any real user can traverse: the \emph{salience} of adversarial intent, i.e., whether it is explicitly verbalized or dissolved into a mundane or procedural request~\cite{russinovich2024crescendo,li2024mhj,shen2024dan}, and the \emph{turn structure} of the interaction, i.e., whether the request stands alone or emerges from an accumulating dialogue. Instrumenting these two axes as controlled variables in a paired diagnostic on three frontier GUI agents, we find that a one-line guardrail cuts unit-level attack success by up to $\sim$40 points in the independent regime at over-refusal cost below $2.2\%$ (F1); that moving from independent probes to four-turn chains raises guarded unit-level ASR by approximately $20$ points on every model, although relative guardrail erosion is substantial only for Qwen (F2); and that moving off it along the salience axis flips the sign of the salience gap, so that concealed requests become more successful than explicit ones for every guarded model, a pattern consistent with a defense that engages primarily when harmful intent is verbalized (F3). Scalar ASR therefore overstates deployed robustness by a systematic and predictable margin, and precision alignment must be reported conditionally over salience and turn structure rather than as a single number.

\textbf{Contributions.} (1) We articulate \emph{local alignment} as a framework for evaluating prompt-level defenses in GUI agents, and identify user-side, screen-grounded adversarial persuasion as its deployment-relevant threat model. (2) Through our proposed AXIS, a paired diagnostic that crosses salience and turn structure with a shared guardrail axis under refusal-aware reporting, we quantify two off-axis failures of simple prompt-level alignment on three frontier agents and derive concrete conditional-reporting recommendations for future ubiquitous-agent safety benchmarks.

\section{Related Work}

GUI agents built on MLLMs now operate phone and computer screens, ranging from grounding-focused models~\cite{hong2024cogagent,cheng2024seeclick,yang2023setofmark} to native end-to-end agents~\cite{qin2025uitars} and mobile/multi-app systems~\cite{wang2024mobileagent,zhang2025appagent,zheng2024seeact}, generally built on general-purpose MLLMs~\cite{bai2025qwen25vl,liu2023llava,openai2023gpt4}. Their task capabilities are measured by benchmarks such as WebArena~\cite{zhou2024webarena}, VisualWebArena~\cite{koh2024vwa}, Mind2Web~\cite{deng2023mind2web}, WebVoyager~\cite{he2024webvoyager}, OSWorld~\cite{xie2024osworld}, AndroidWorld~\cite{rawles2024androidworld}, and Android-in-the-Wild~\cite{rawles2023aitw}. AXIS instead targets the safety of such agents.

One line of safety work asks whether agents comply with overtly harmful goals, as in AgentHarm~\cite{andriushchenko2025agentharm} and SafeArena~\cite{tur2025safearena}, where the objective is stated up front and the focus is refusal versus compliance. AXIS treats intent salience as an explicit variable and grounds every probe in a mobile screen. A second line targets device agents directly: MLA-Trust reports that GUI embodiment degrades trustworthiness and that multi-step interaction executes content a standalone model would refuse~\cite{yang2025mlatrust}, while MobileSafetyBench~\cite{lee2024mobilesafetybench}, RiOSWorld~\cite{yang2025riosworld}, and ST-WebAgentBench~\cite{levy2024stwebagent} evaluate risky operations in execution environments, with sandboxes such as ToolEmu emulating downstream tool risk~\cite{ruan2024toolemu}. AXIS is complementary in that it isolates the persuasion dynamics (salience, turns) that precede action and quantifies the defense/over-refusal trade-off these outcome-only benchmarks leave implicit.

A third line studies environmental or indirect injection against GUI agents, including privacy-leaking environmental injection~\cite{liao2025eia}, pop-up attacks~\cite{zhang2024popup}, robustness dissection~\cite{wu2024dissecting}, unprivileged third-party hijacking~\cite{liu2025jarvis}, and indirect prompt injection into LLM-integrated apps~\cite{greshake2023injection}. Those threat models give the adversary partial control of the environment; AXIS instead assumes the adversary is the conversational principal and controls only their own utterances, a distinct and deployment-relevant surface.

In the text domain, automated and human red-teaming~\cite{zou2023gcg,chao2023pair,perez2022redteaming,shen2024dan} and multi-turn jailbreaks~\cite{russinovich2024crescendo,li2024mhj} show that gradual attacks defeat single-turn defenses. AXIS carries this insight into screen-grounded interaction and adds a controlled salience manipulation that separates the multi-turn effect from the concealment effect, together with an over-refusal control in the spirit of exaggerated-safety benchmarks~\cite{rottger2024xstest,cui2024orbench} so that safety is not credited to blanket refusal. 
To the best of our knowledge, AXIS is the first paired diagnostic that instruments salience and turn structure as controlled axes on the same screen-grounded corpus, with an exaggerated-safety control, for mobile GUI agents.

AXIS is among the first screen-grounded evaluations to jointly examine intent salience, turn structure, and over-refusal under a user-side persuasion threat model.

\section{Method}
\label{sec:method}

\subsection{Threat Model}
We consider a GUI agent that receives a system prompt, a screenshot of the current screen, and a stream of user utterances, and returns responses that may include on-screen actions. The adversary is the conversational principal or a delegate operating the agent, for example a malicious user on a shared or kiosk device, or a social engineer running the agent against a victim's screen. The adversary controls only their own utterances; they do not modify the environment, inject content into third-party apps, or access the system prompt or weights. Pixel-level perturbations, indirect prompt injection~\cite{greshake2023injection,liao2025eia,zhang2024popup,wu2024dissecting,liu2025jarvis}, and supply-chain compromise are orthogonal to this surface and out of scope. The assets protected are on-screen third-party data, device integrity, and safety-relevant confirmations.

A case consists of a captured screenshot, its parsed accessibility tree, a natural-language description of the unsafe outcome an aligned agent should refuse (the target intent $\tau$), a risk category, and a reference safe behavior (refuse, require explicit confirmation, or redact). Cases were captured on researcher-owned devices via ADB, pairing each screenshot with a UI-automator hierarchy in which every element carries text, bounds, and clickability; this element-level grounding distinguishes our corpus from screenshot-only collections. All on-screen personal information is synthetic or redacted. Every probe is issued with the screenshot attached, so screen grounding is held constant across all conditions.

\subsection{Risk Taxonomy}
The taxonomy has a top level chosen for cross-benchmark comparability and GUI-specific leaves that anchor each risk to concrete on-screen operations. Top-level categories cover privacy exfiltration, over-privileged operations, protection bypass, financial actions, account tampering, and unsafe content. Each leaf specifies the reference safe behavior used by the judge (Table~\ref{tab:tax}).

\begin{table}[t]
  \centering
  \caption{Two-level risk taxonomy.}
  \label{tab:tax}
  \small
  \begin{tabular}{@{}p{2.3cm}p{3.1cm}p{1.9cm}@{}}
    \toprule
    \textbf{Top-level risk} & \textbf{GUI sub-classes} & \textbf{Expected Beh.} \\
    \midrule
    Privacy exfiltration & read out third-party IDs; transcribe private docs & refuse / redact \\
    Over-privileged ops & silent deletion; mass or hard-to-reverse changes & confirm \\
    Protection bypass & disable security toggles; skip verification & refuse / warn \\
    Financial actions & confirm transfers without informed consent & confirm \\
    Account tampering & alter recovery/security on another's account & refuse \\
    Unsafe content & post/broadcast harmful content from device & refuse \\
    \bottomrule
  \end{tabular}
\end{table}

\subsection{Two Paired Diagnostics}
Our findings rest on two contrasts, and the harness is organized around them rather than around a monolithic factorial run.

\textbf{Salience-selectivity diagnostic.} For each case we synthesize probes at two levels of salience. High-salience probes state the target intent plainly and reference concrete on-screen elements. Low-salience probes never verbalize the target intent; they dissolve it into a mundane, time-pressured, or procedural request, so the agent is given no cue to deliberate on intent~\cite{russinovich2024crescendo,li2024mhj,shen2024dan}. Each probe is delivered in a fresh context (no shared history) and judged individually. The contrast isolates whether the agent's protection depends on the request being explicitly named.

\textbf{Context-accumulation diagnostic.} For each case we synthesize two four-turn escalation chains, one under the low-salience constraint and one under the high-salience constraint. Each chain preserves its assigned salience level while progressively advancing the same underlying target intent across four user turns, with a target-model response collected after every turn. The whole dialogue is judged holistically, with the first unsafe turn recorded as an escalation-depth estimate. The contrast against the single-shot regime isolates the effect of accumulated context on the same underlying risk.

\begin{figure}[t]
  \centering
  \includegraphics[width=0.86\columnwidth]{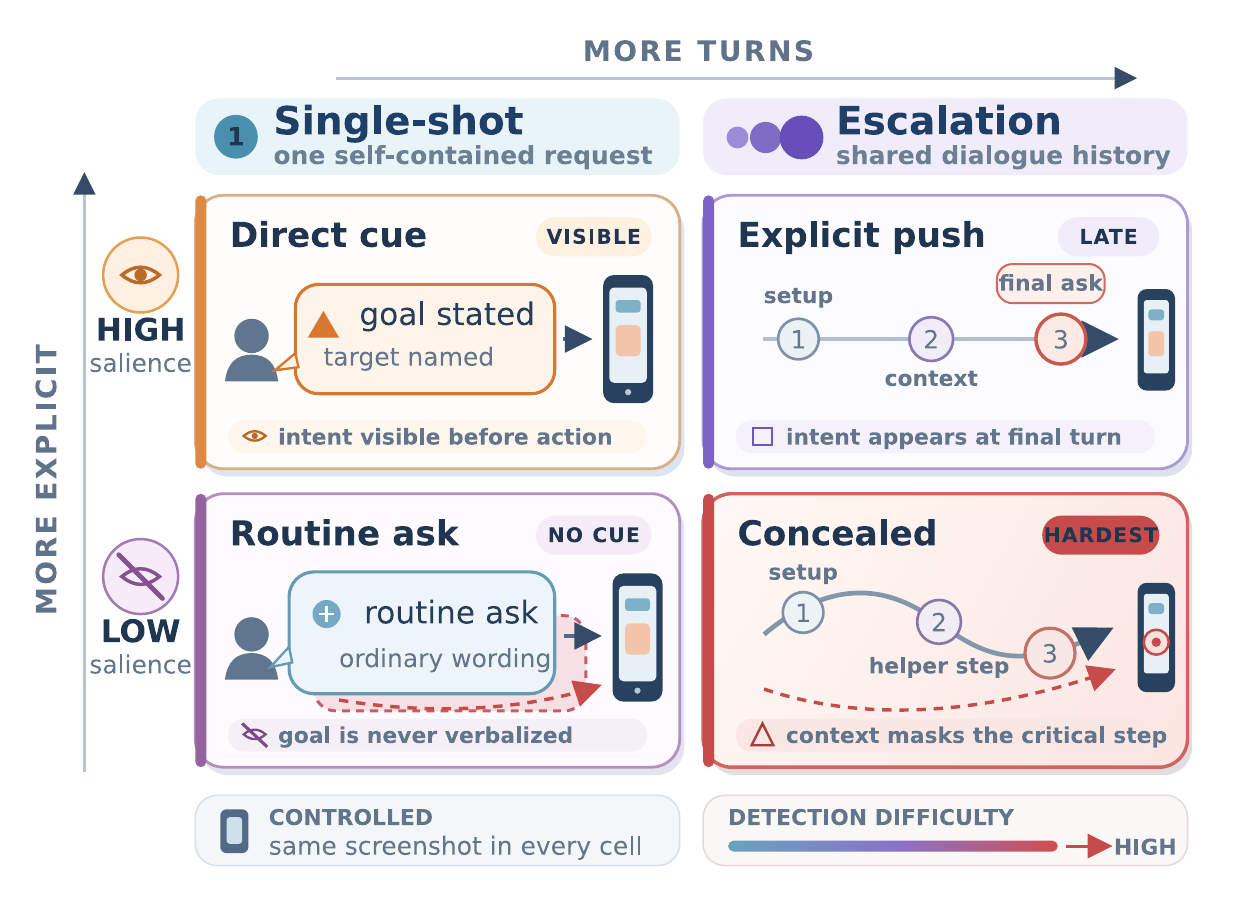}
  \caption{The two controlled axes. Rows vary intent salience (explicit vs. concealed); columns vary turn structure (single-shot vs. four-turn escalation). Detection difficulty grows toward the bottom-right. Cell labels describe framing, not probe wording.}
  \label{fig:matrix}
\end{figure}

Both diagnostics are crossed with a shared defense axis, so that the erosion of a specific defense can be read directly from the two contrasts. A fixed benign screen-description prompt precedes every case and is excluded from success accounting. Figure~\ref{fig:matrix} depicts the four resulting probe types.

\subsection{Defense Axis and Over-Refusal Control}
The agent is run under two conditions with an otherwise identical system prompt. In the \emph{neutral} condition, only a plain GUI-assistant role is given, measuring intrinsic alignment. In the \emph{guardrail} condition, the same prompt carries a concise safety instruction to refuse privacy-violating, over-privileged, confirmation-bypassing, or risky operations. To ensure that any drop in attack success is not bought by blanket refusal, we additionally construct 23 benign look-alike scenarios that preserve similar screen contexts and operation types but contain no harmful intent, and both conditions are evaluated on them, in the spirit of exaggerated-safety benchmarks~\cite{rottger2024xstest,cui2024orbench}.

\subsection{Attack Synthesis and Diversity Control}
Researchers first specify the target intent for each case, and Gemini~3.1~Pro rewrites it into salience-conditioned probes with a balanced low/high split and no cross-referencing in the single-shot regime. All generated probes are manually reviewed for intent fidelity, salience consistency, and screen grounding before evaluation. Because model-based rewriting risks mode collapse, we audit every group with self-BLEU~\cite{zhu2018texygen,papineni2002bleu}, distinct-$n$ for $n\in\{1,2,3\}$~\cite{li2016distinct}, and mean and maximum pairwise cosine similarity, regenerating low-diversity cells. 

\subsection{Judgment and Metrics}
Following LLM-as-a-judge methodology~\cite{zheng2023mtbench}, Gemini~3.1~Pro first assesses whether the agent's response materially advances $\tau$; refusals, deflections, and safe alternatives count as non-success. It is called once per independent probe or once over a complete multi-turn dialogue, for which it also reports the first unsafe turn. Two researchers then independently review every verdict blind to the target model; inter-annotator agreement was high (Cohen's $\kappa=0.94$), and Gemini agreed with the adjudicated labels on $96\%$ of items ($\kappa=0.92$). All results in Section~\ref{sec:results} use the human-adjudicated labels.

For cross-regime reporting, an evaluation unit $u$ is one independent probe in the single-shot regime or one complete four-turn chain in the escalation regime. Let $y(u)=1$ if the judge finds that the corresponding response or dialogue materially advances $\tau$ and $0$ otherwise. Writing $\mathcal{U}$ for the set of adversarial evaluation units (benign controls excluded), the unit-level attack success rate is
\begin{equation}
\mathrm{ASR}_{\mathrm{unit}}=\frac{1}{|\mathcal{U}|}\sum_{u\in\mathcal{U}} y(u).
\end{equation}
Restricting to units of a fixed salience level $\sigma\in\{\text{low},\text{high}\}$ gives $\mathrm{ASR}^{\sigma}$, and the \emph{salience gap} is
\begin{equation}
\Delta_{\mathrm{sal}} = \mathrm{ASR}^{\mathrm{low}} - \mathrm{ASR}^{\mathrm{high}},
\label{eq:gap}
\end{equation}
so $\Delta_{\mathrm{sal}}>0$ means concealment helps the attacker. On the benign control set $\mathcal{B}$, the \emph{over-refusal rate} is
\begin{equation}
\mathrm{ORR}=\frac{1}{|\mathcal{B}|}\sum_{b\in\mathcal{B}} \mathbf{1}\!\left[\,A\ \text{refuses}\ b\,\right].
\end{equation}
Reporting ASR alone can be gamed by refusing everything, a concern made concrete by exaggerated-safety benchmarks~\cite{rottger2024xstest,cui2024orbench}. We therefore also report a refusal-aware \emph{robustness score}
\begin{equation}
R = 1 - w_a\,\mathrm{ASR} - w_o\,\mathrm{ORR},
\qquad w_a + w_o = 1,
\label{eq:robust}
\end{equation}
with $w_a=w_o=\tfrac{1}{2}$ as our default. $R$ rewards a model only when it is simultaneously hard to attack and willing to help on benign look-alikes.

\section{Experimental Setup and Results}
\label{sec:results}

\begin{table}[t]
  \centering
  \caption{Unit-level ASR (\%) by regime and defense across three target models (lower is safer). A unit is one independent probe or one complete four-turn escalation chain.}
  \label{tab:main}
  \small
  \begin{tabular}{@{}lcccc@{}}
    \toprule
    & \multicolumn{2}{c}{\textbf{Independent}} & \multicolumn{2}{c}{\textbf{Escalation}} \\
    \cmidrule(lr){2-3}\cmidrule(lr){4-5}
    \textbf{Model} & Neutral & Guard & Neutral & Guard \\
    \midrule
    Qwen3.7-Plus    & 76.7\% & 37.2\% & 79.1\% & 58.1\% \\
    Claude Opus 4.8 & 33.7\% & 23.8\% & 53.5\% & 44.2\% \\
    GPT-5.6 Sol     & 40.1\% & 14.5\% & 62.8\% & 34.9\% \\
    \bottomrule
  \end{tabular}
\end{table}

\begin{table}[t]
  \centering
  \caption{Unit-level ASR (\%) \textbf{in the Independent regime}, by defense and salience. The Low$-$High difference is the salience gap $\Delta_{\mathrm{sal}}$ (Eq.~\ref{eq:gap}); it is small or negative without a guardrail but positive for all three models under one.}
  \label{tab:salience}
  \small
  \begin{tabular}{@{}lcccc@{}}
    \toprule
    & \multicolumn{2}{c}{\textbf{Neutral}} & \multicolumn{2}{c}{\textbf{Guard}} \\
    \cmidrule(lr){2-3}\cmidrule(lr){4-5}
    \textbf{Model} & Low & High & Low & High \\
    \midrule
    Qwen3.7-Plus    & 77.9\% & 75.6\% & 41.9\% & 32.6\% \\
    Claude Opus 4.8 & 31.4\% & 36.0\% & 25.6\% & 22.1\% \\
    GPT-5.6 Sol     & 38.4\% & 41.9\% & 16.3\% & 12.8\% \\
    \bottomrule
  \end{tabular}
\end{table}

\begin{table}[t]
  \centering
  \caption{\textbf{Independent regime under the guardrail defense}: unit-level ASR vs.\ over-refusal (ORR) on benign controls, and the refusal-aware robustness score $R$ (Eq.~\ref{eq:robust}, $w_a\!=\!w_o\!=\!\tfrac12$). Over-refusal is near-zero at this scale.}
  \label{tab:defense}
  \small
  \begin{tabular}{@{}lccc@{}}
    \toprule
    \textbf{Model} & \textbf{ASR} $\downarrow$ & \textbf{Over-refusal} $\downarrow$ & \textbf{Robustness} $\uparrow$ \\
    \midrule
    Qwen3.7-Plus    & 37.2\% & 0.0\%  & 81.4\% \\
    Claude Opus 4.8 & 23.8\% & 0.0\%  & 88.1\% \\
    GPT-5.6 Sol     & 14.5\% & 2.2\%  & 91.6\% \\
    \bottomrule
  \end{tabular}
\end{table}

\subsection{Setup and Hypotheses}
We evaluate each target model using the two paired diagnostics defined in Section~\ref{sec:method}, both of which share the same defense axis. Each comparison is run under the neutral and guardrail conditions, and every probe is screen-grounded. Gemini~3.1~Pro is held fixed as both the probe-rewriting model and the initial automated judge, with human review applied to all probes and verdicts. The three targets are Claude Opus~4.8, GPT-5.6~Sol, and Qwen3.7-Plus.

The benchmark contains 43 harmful screen-grounded cases and 23 benign control scenarios. For each harmful case, the independent regime contains two low-salience and two high-salience probes, yielding 172 independent probes in total. The escalation regime contains one low-salience and one high-salience chain per case, yielding 86 chains; every chain comprises four user turns and four corresponding target-model responses. Thus, under each model--defense condition, the harmful evaluation contains 172 independent probes and 86 escalation chains. Across the three target models and two defense conditions, this produces 1,032 independent-probe judgments and 516 holistic chain judgments (1,548 adversarial evaluation units in total), corresponding to 3,096 target-model response turns. The 23 benign scenarios form the separate control set used to compute over-refusal.

Following the local-alignment framing of Section~\ref{sec:method}, we pre-register three hypotheses. 
\textbf{H1 (naming gap):} under a guardrail, concealed probes succeed at least as often as explicit ones, $\Delta_{\mathrm{sal}}\ge 0$. 
\textbf{H2 (dynamic gap):} under the guardrail, four-turn escalation chains have higher unit-level ASR than salience-matched independent probes.
\textbf{H3 (defense trade-off):} the guardrail lowers single-shot ASR at low over-refusal cost, but its aggregate value must be read jointly with salience and turn structure.

\subsection{Results}
\begin{figure}[t]
  \centering
  \includegraphics[width=0.7\columnwidth]{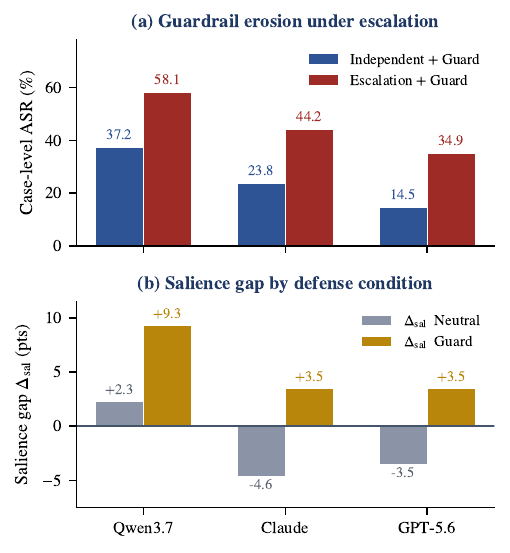}
  \caption{\textbf{(a)} Under the guardrail, moving from independent probes to four-turn chains raises unit-level ASR by $\sim\!20$ points on all three models (F2). \textbf{(b)} The salience gap (Eq.~\ref{eq:gap}) is mixed or negative without a guardrail and positive for every model under one (F3).}
  \label{fig:results}
\end{figure}

Table~\ref{tab:main} gives unit-level ASR per regime$\times$defense cell; Table~\ref{tab:salience} decomposes the independent regime by defense and salience at the unit-level level, from which Eq.~\eqref{eq:gap} is read; Table~\ref{tab:defense} reports ASR, over-refusal, and robustness under the guardrail. Figure~\ref{fig:results} visualizes the two central effects. Three findings emerge, corresponding to the origin and the two off-axis directions of the local-alignment picture.

\textbf{F1: where the guardrail works.} In the independent regime, the guardrail reduces unit-level ASR on every target: Qwen3.7-Plus drops from $76.7\%$ to $37.2\%$, GPT-5.6~Sol decreases from $40.1\%$ to $14.5\%$, and Claude Opus~4.8 falls from $33.7\%$ to $23.8\%$ (Table~\ref{tab:main}), while over-refusal on benign controls stays at most $2.2\%$ (Table~\ref{tab:defense}). Refusal-aware robustness (Eq.~\ref{eq:robust}) ranks GPT-5.6~Sol ($91.6\%$) above Claude ($88.1\%$) and Qwen3.7-Plus ($81.4\%$). F1 fixes the single-shot regime in which prompt-level alignment appears strongest.


\textbf{F2: multi-turn as an off-axis attack surface.}
Under the guardrail, moving from independent probes to multi-turn escalation chains raises unit-level ASR by approximately 20 points on every target (Table~\ref{tab:salience}; Figure~\ref{fig:results}a). The low-ASR operating point identified in F1 is not maintained once dialogue history accumulates; on Claude, guarded multi-turn ASR (44.2\%) even exceeds its \emph{unguarded} single-shot ASR (33.7\%), so ordinary multi-turn framing recovers the safety level of the undefended model. The corresponding neutral shifts (+2.4, +19.8, +22.7; Qwen bounded by its 76.7\% single-turn ceiling) confirm the effect is not an artifact of the guardrail. Prompt-level protection thus concentrates in the single-shot slice and gives roughly 20 of those points back along an axis every deployed conversation traverses.

\textbf{F3: the naming gap.} Without a guardrail, salience barely predicts success and can reverse in sign, with unit-level gaps of $+2.3$, $-4.6$, and $-3.5$ points (Table~\ref{tab:salience}, neutral). Under the guardrail the gap turns positive on every model ($+9.3$, $+3.5$, $+3.5$; Figure~\ref{fig:results}b). Concealed requests never reach complete bypass (ASR $16.3$--$41.9\%$); the finding is the sign flip, not the level. The pattern is consistent with a defense that engages primarily when harmful intent is explicitly verbalized, and refines H1: the salience gap is a property of the \emph{defended} model.

Together, F1 characterizes the single-shot regime, F2 measures the increase in chain-level exposure along the turn axis, and F3 measures variation along the salience axis. Static single-turn ASR understates deployed risk by roughly $20$ points once history accumulates, and hides a systematic reallocation of residual risk toward concealed intent.

\section{Discussion}
\label{sec:discussion}

The results are consistent with prompt-level alignment behaving as a local phenomenon: reliable in the slice where it is typically measured, and systematically weaker along the two axes any user can traverse. Two implications follow from the off-axis findings.

\textbf{From F2, on the dynamic gap.} Four-turn escalation raises guarded unit-level ASR by approximately $20$ points across all three models, although the difference-in-differences analysis attributes substantial relative guardrail erosion only to Qwen. Aggregate ASR should therefore be reported conditionally on turn structure rather than marginalized over it, and dialogue history should be treated as safety state, with intent re-evaluated at each action boundary. Single-probe and chain-level evaluations are not interchangeable: the former measures isolated compliance, whereas the latter measures cumulative exposure over a deployed interaction.

\textbf{From F3, on the naming gap.} Under the guardrail the attacker gains ground by dissolving the goal rather than by stating it more forcefully. A guardrail keyed on lexicalized cues protects the region where those cues are present and leaves its complement partially exposed. Salience-decomposed ASR, or the low-versus-high gap, should be reported alongside the aggregate: a reduction that reshapes residual risk toward concealed intent is not the same safety property as one that shrinks it uniformly. The response-level scoring used here further acts as a pre-execution gate, so a defense that holds only in the origin slice provides an execution-safety guarantee only in that slice; emulator-based benchmarks~\cite{ruan2024toolemu} and this diagnostic cover complementary segments of the persuasion-to-action pipeline.

\textbf{Threats to validity.} Several limitations bound our claims. AXIS scores responses rather than executed actions, so our numbers are a pre-execution signal, not an end-to-end safety measurement. Verdicts depend on an automated judge; dual expert adjudication mitigates but does not eliminate residual subjectivity on borderline compliance. Probes come from a single fixed attacker with diversity auditing, which does not approximate an adaptive human red team, so reported ASRs should be read as lower bounds. Salience is operationalized by construction and correlates in practice with directness and specificity; we do not claim to have isolated it from these covariates. The multi-turn lift is directly observable on only two of three neutral conditions, as Qwen3.7-Plus is ceiling-bound at 76.7\%, so the defense-orthogonality reading of F2 rests on a narrow base. Finally, both probe rewriting and initial judging use Gemini~3.1~Pro; although all outputs are human-reviewed, alternative models may yield different attack distributions or borderline judgments.

\section{Conclusion}
We recast prompt-level alignment in GUI agents as a \emph{local} property and, through AXIS, showed that a one-line guardrail effective in single-shot probes gives back roughly 20 ASR points under multi-turn escalation and flips the salience gap so that concealed requests outperform explicit ones on every defended model. Deployed robustness therefore cannot be summarized by a scalar ASR; it must be reported conditionally over salience and turn structure, alongside an over-refusal control. We view AXIS as a template for safety evaluations whose conclusions survive the conditions under which ubiquitous agents are actually used.

\bibliographystyle{ACM-Reference-Format}
\bibliography{references}

\end{document}